\documentclass[aps,twocolumn,a4paper,showpacs]{revtex4}
\usepackage{graphicx}
\usepackage{amsmath}
\usepackage{amssymb}
\usepackage{enumerate}
\usepackage{subfigure}
\usepackage{tabularx}
\newcommand{\be}{\begin{equation}}
\newcommand{\ee}{\end{equation}}
\newcommand{\ben}{\begin{eqnarray}}
\newcommand{\een}{\end{eqnarray}}
\newcommand{\bes}{\begin{subequations}}
\newcommand{\ees}{\end{subequations}}

\newcommand{\bb}{\bibitem}
\newcommand{\sech}{{\rm sech}}

\begin{document}
\title{From Scalar Field Theories to Supersymmetric Quantum Mechanics}
\author{D. Bazeia$^{1}$ and F.S. Bemfica$^{2}$}
\affiliation{$^1$Departamento de F\'\i sica, Universidade Federal da Para\'\i ba, 58051-970 Jo\~ao Pessoa, PB, Brazil}
\affiliation{$^2$Escola de Ci\^encias e Tecnologia, Universidade Federal do Rio Grande do Norte, 59072-970 Natal, RN, Brazil}
\begin{abstract}
In this work we report a new result that appears when one investigates the route that starts from a scalar field theory and ends on a supersymmetric quantum mechanics. The subject has been studied before in several distinct ways and here we unveil an interesting novelty, showing that the same scalar field model may describe distinct quantum mechanical problems.
\end{abstract}
\date{\today}
\maketitle

\section{Introduction}

The study of topological structures in high energy physics was initiated long ago, and a diversity of important contributions to the subject can be found, for instance, in Refs.~\cite{b1,b2,b3,b4,b5}. As a particular fact, one knows that among the several possibilities to construct localized structures of the kink, vortex, monopole and other types, the case of kinks seems to be the simplest one, since it appears in $(1,1)$ spacetime dimensions, and may be governed by a single real scalar field. An interesting aspect of kinks is that the calculation concerning linear stability results in a special kind of quantum physics, known as supersymmetric quantum mechanics \cite{fac,susy}.

Previous investigations on the subject that we are interested in the current work appeared before, for instance, in Refs.~\cite{pw1,pw2,pw3,pw4,pw5}. In particular, in \cite{pw4,pw5} the authors studied the reconstruction of the field theory model from reflectionless scattering data. An interesting result there suggested is that for reflectionless potentials, the field theory model seems to be like the sine-Gordon model when there is only one bound state, the zero or translational mode, or like the $\phi^4$ model when two bound states are present and a specific choice of the eigenvalues is taken.

In this work we revisit the problem to report a strong result, shedding further light on the subject. We follow the standard route, but study models with distinct topological sectors, and elaborate a general result, which follows after describing two well distinct examples, one relying on polynomial potentials of the $\phi^4$ type, and the other on periodic potentials of the sine-Gordon type. The main result shows that a scalar field theory that engenders distinct topological sectors may lead to distinct stability potentials that can be used to describe distinct supersymmetric quantum mechanical problems.

We organize the work as follows. In Sec.~\ref{sec2} we review the subject and introduce the basic results already known in the literature. We turn attention to the case of scalar field models that engenders distinct topological sectors in Sec~\ref{sec3}, where we investigate specific models and then add a general statement, which we prove at the very end of section. We end the work in Sec.~\ref{sec4}, where we summarize the results and comment on specific issues under current consideration, to be reported in the near future.

\section{Illustration}
\label{sec2}

The work is based on $(1,1)$ spacetime dimensions, and deals with a single real scalar field $\phi$. The models are described by the Lagrange density
\be\label{mod}
{\cal L} = \frac12 \partial_\mu\phi \partial^\mu \phi - V(\phi),
\ee
where $\phi$ is the field and
\be\label{pot}
V(\phi) = \frac12 W^2_\phi,
\ee
is the potential, with $W=W(\phi)$ being a real function of the scalar field, such that
\be\label{sup}
W_\phi=\frac{d W}{d\phi}.
\ee
This is a nice way to describe the model, since the potential is now explicitly non-negative, so it is limited from below, with the absolute minima obeying $W_\phi=0$. We consider models that support a countable set of minima, supposing that $W_\phi(v_ i)=0$ for $i=1,2,\cdots$. The field $\phi$, the space $x$ and the time $t$ are all redefined, in a way such that they are all dimensionless, so the work is written using dimensionless quantities.

The equation of motion is given by
\be\label{eom}
\frac{\partial^2\phi}{\partial t^2}-\frac{\partial^2\phi}{\partial x^2}+W_\phi W_{\phi\phi}=0,
\ee
and in the case of a static configuration one gets
\be\label{eoms}
\frac{d^2\phi}{dx^2}=W_\phi W_{\phi\phi}.
\ee
Here we introduce the first-order differential equations
\be\label{foe}
\frac{d\phi}{dx}=\pm W_\phi
\ee
and one sees that solutions of \eqref{foe} also obey the equation of motion \eqref{eoms}. The two signs are used to describe kinks and antikinks.

As one knows, two adjacent minima of the potential define a topological sector, and for solutions that obey the first-order equations \eqref{foe}, the energy has the form
\be
E_k=|W(v_k)-W(v_{k+1})|.
\ee
The energy density $\rho_k(x)$ can be written as
\ben
\rho_k(x)=\left(\frac{d\phi_k}{dx}\right)^2,
\een
where $\phi_k(x)$ is static solution of \eqref{foe}, which behaves as $\phi_k(x\to-\infty)\to v_k$ and $\phi_k(x\to\infty)\to v_{k+1}$, if it is a kink, or $\phi_k(x\to-\infty)\to v_{k+1}$ and $\phi_k(x\to\infty)\to v_k$, if it is an antikink.

To study linear stability, for simplicity we omit the label that identify the topological sector under investigation. We suppose that the model contains at least one topological sector with static solution $\phi(x)$, and the fluctuations are included in the form
\be\label{exp}
\phi(x,t)=\phi(x)+\sum_n \eta_n(x)\cos(w_n t).
\ee
After substituting this into the equation of motion \eqref{eom} one writes, up to first-order in the fluctuations,
\be\label{qm}
\left(-\frac{d^2}{dx^2}+U(x)\right)\eta_n(x)=w^2_n \eta_n(x),
\ee
where $U(x)$ is the stability potential. It has the form
\be\label{U}
U(x)=W^2_{\phi\phi}+W_\phi W_{\phi\phi\phi},
\ee
and must be calculated at the classical solution $\phi(x)$ used in \eqref{exp}. It is interesting to see that the above second-order operator can be factorized in the form
\be
-\frac{d^2}{dx^2}+U(x)= S^\dag S,
\ee
where
\be
S=\frac{d}{dx}+f(x),
\ee
and
\be\label{f}
f(x)= \mp W_{\phi\phi}.
\ee
Here the upper or lower sign has to be chosen properly, following the choice of the upper or lower sign in \eqref{foe} and the choice of $W$ itself.

The procedure requires that the static solution solves the first-order equation \eqref{foe}. Also, because the operator
$S^\dag S$ is non-negative, one is led to the conclusion that the static solution is linearly stable. Moreover, since the scalar field model \eqref{mod} engenders translational invariance, the normalized zero mode $S\eta_t(x)=0$ has to be present. It can be written as
\be
\eta_t(x)=a \exp\left(-\int dx\, f(x)\right)=\frac{N}{w(x)},
\ee
where $N$ is the normalization constant, and $w(x)$ is such that
\be\label{fw}
f(x)=\frac{w'(x)}{w(x)},
\ee
where the prime stands for derivative of the function with respect to its argument, that is, $w'(x)=dw/dx$.

As it is well-known, one can write the zero mode as the derivative of the kinklike solution itself. The proof follows directly from the equation of motion for the static field \eqref{eoms}, and one can then write $\eta_t(x)\propto\phi'(x)= {1}/{w(x)}$, and this allows that we write
\be
V(\phi)=\frac12 \left(\frac{1}{w^2(x)}\right)_{x=x(\phi)}.
\ee
In the above result one has to use $x=x(\phi)$, that is, one needs to invert the kinklike solution $\phi=\phi(x)$; when this is done analytically, one gets the potential analytically.

The above calculations follow naturally from the standard results, and if one uses \eqref{U} and \eqref{f} it is possible to write the stability potential in the form
\be\label{Uf}
U(x)=f^2(x)-f^\prime(x).
\ee
Alternatively, one can use \eqref{fw} to get
\be
U(x)=2\frac{w^{\prime2}(x)}{w^2(x)}-\frac{w^{\prime\prime}(x)}{w(x)}.
\ee
The procedure unveils a direct route from field theory to quantum mechanics. In fact, it leads to supersymmetric quantum mechanics, since it is always possible to include another non-negative operator, $S S^\dag$, which is the supersymmetric partner of $S^\dag S$; see, e.g., Ref.~{\cite{susy}}.

We note that the study of linear stability leads to an associated quantum mechanical problem, described by the Schr\"odinger-like equation that appears in \eqref{qm}. One then sees that a scaling in the $x$ coordinate in the form $x\to\alpha x$, leads to an equivalent problem with the modification $U(x)\to\alpha^2U(\alpha x)$. This observation will be useful in the investigation that follows below.

To illustrate how a scalar field theory drives us to quantum mechanics, let us consider the Higgs prototype, the $\phi^4$ model with spontaneous symmetry breaking. The model is described by the potential
\be
V=\frac12(1-\phi^2)^2.
\ee
Here we write
\be
W=\phi-\frac13 \phi^3,
\ee
and the first-order equations become
\be
\phi^\prime =\pm1\mp\phi^2.
\ee
The two minima $v_\pm=\pm1$ define the topological sector that supports the kink $\phi(x)=\tanh(x)$ and the antikink $\phi(x)=-\tanh(x)$, both with the same energy $E=4/3$. We use \eqref{f} to get
\be
f(x)=2\tanh(x),
\ee
and now Eq.~\eqref{Uf} leads to the stability potential
\be
U(x)=4-6\,\sech^2(x),
\ee
which is the well-known modified P\"oschl-Teller potential \cite{PT,MF}. As one knows, it supports two bound states, the zero mode and a massive state.

We can investigate another model, the $\phi^6$ model described by \cite{lohe}
\be
V=\frac12\phi^2(1-\phi^2)^2.
\ee
Here we can write
\be
W=\frac12\phi^2-\frac14\phi^4,
\ee
and the first-order equations
\be
\phi^\prime =\pm\phi(1-\phi^2).
\ee
There are three minima, $v_0=0$ and $v_\pm=\pm1$, and now the model has two topological sectors that are equivalent to each other. They have the same energy $E=1/4$. The kinks and antikinks are given by
\be
\phi(x)=\pm \left(\frac12(1\pm\tanh(x))\right)^{1/2}.
\ee
Here we can write
\be
f(x)=\frac12+\frac32\tanh(x)
\ee
and the stability potential has the form
\be
U(x)=\frac52-\frac{15}{4}\sech^2(x)+\frac32\tanh(x).
\ee
This potential supports the zero mode and no other bound state, but there is another stability potential, which is $U(-x)$; it is the scaled potential $\alpha^2 U(\alpha x)$, with $\alpha=-1$, so $U(x)$ and $U(-x)$ represent equivalent quantum mechanical problems.

Another illustration of interest concerns the sine-Gordon model. Here we consider the potential
\be\label{sg}
V=\frac12 \cos^2(2\phi),
\ee
and now
\be
W=\pm\frac12\,\sin(2\phi).
\ee
The first-order equation for the kinklike solution is
\be
\phi^\prime=\cos(2\phi),
\ee
This case is different since the model has an infinite number of minima $v_k=k\pi/4$, for $k=\pm1,\pm3,\pm5,\cdots$. It shows that the model has an enumerable set of topological sectors, but they are all equivalent, having the same energy $E=1$. Thus, we focus on the basic solution that lives in the central sector, with $-\pi/4\leq \phi\leq\pi/4$. It has the form
\be
\phi(x)=\frac12\arcsin(\tanh(2x)).
\ee
In this case one gets
\be
f(x)=2\tanh(2x),
\ee
and the stability potential becomes
\be\label{psg}
U(x)=4-8\,\sech^2(2x).
\ee
We note that the scaling $x\to x/2$ allows to write this potential in the standard form $U(x)=1-2 \,\sech^2(x)$. As one knows, it supports the zero model and no other bound state.

\section{New results}
\label{sec3}

The above calculations appear in several investigations and are nicely studied in the two works \cite{pw4,pw5}, but now we focus on bringing novelty to the subject, which arises when the scalar field model engenders distinct topological sectors. For simplicity, we concentrate on models that support two distinct topological sectors, searching for the respective stability potentials.

\begin{figure}[t]
\includegraphics[width=7cm]{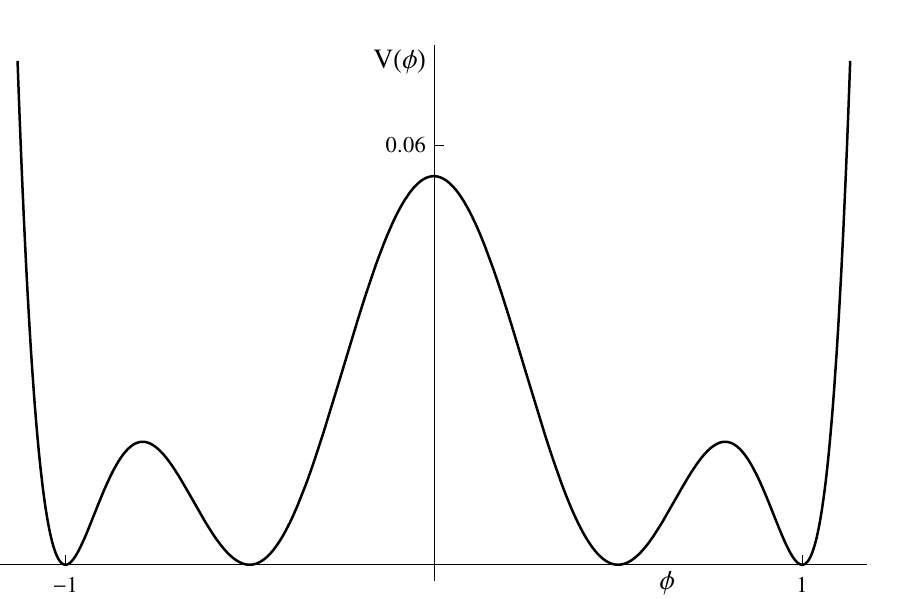}
\caption{The $\phi^8$ model \eqref{pot8} and the three topological sectors.}\label{fig1}
\end{figure}

We first consider a polynomial potential of the $\phi^8$ type. The model is
\be\label{pot8}
V(\phi)=\frac89 \left(\frac14-\phi^2\right)^2 (1-\phi^2)^2.
\ee
It is displayed in Fig.~\ref{fig1}. This model was studied before in \cite{DB2}; see also \cite{khare}. The first-order equation is
\be
\phi^\prime=\frac{4}{3}\left(\frac14-\phi^2\right)(1-\phi^2)
\ee
The minima are $\pm1$ and $\pm1/2$. There are three topological sectors, the central one, connecting the minima $\pm1/2$, and the lateral ones, connecting $-1$ to $-1/2$ and $1/2$ to $1$. There are kinklike solutions of the form
\be\label{sol8}
\phi_k(x)=\cos\left(\frac{\Theta_k(x)}{3}\right),
\ee
where we are considering $\Theta_k(x)=\theta(x)+(3-k)\pi$, with $\theta(x)=\arccos(\tanh(x))\in [0,\pi]$ and $k=1,2,3$. One notes that $\phi_1(x)=-\phi_3(-x)$, and these two topological sectors are similar sectors.

We use the solutions \eqref{sol8} to get
\be
w_k(x)=3\cosh(x)\csc\left(\frac{\Theta_k(x)}{3}\right),
\ee
and the potentials
\be\label{p8}
U_k(x)=1-\frac{19}{9}\sech^2(x)+\sech(x)\tanh(x)\cot\left(\frac{\Theta_k(x)}{3}\right).
\ee
One notes that $U_3(x)=U_1(-x)$, so they are equivalent, related to each other via the scaling $U_3(x)=\alpha^2 U_1(\alpha x)$ with $\alpha=-1$; as we will show below, this is related to the fact that $w_3(x)=w_1(-x)$.

\begin{figure}[t]
\includegraphics[width=7cm]{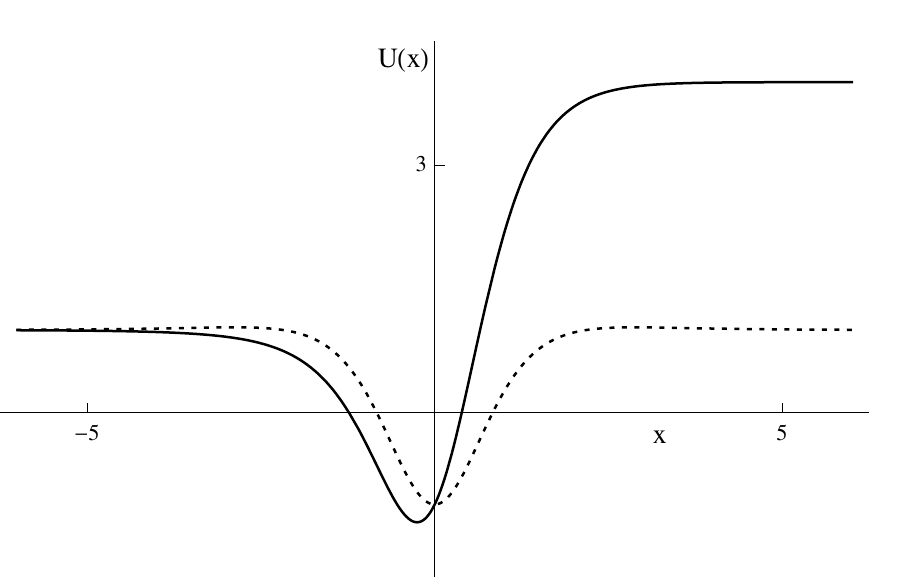}
\caption{The two potentials $U_1(x)$ and $U_2(x)$ that appear in \eqref{p8}, displayed with solid and dotted lines, respectively.}\label{fig2}
\end{figure}

In Fig.~\ref{fig2} we display the two potentials $U_1(x)$ and $U_2(x)$ given by \eqref{p8} to show how they behave. They support the zero mode and no other bound state. Anyway, the two distinct topological sectors present in the scalar field model generate two distinct stability potentials.

\begin{figure}[t]
\includegraphics[width=7cm]{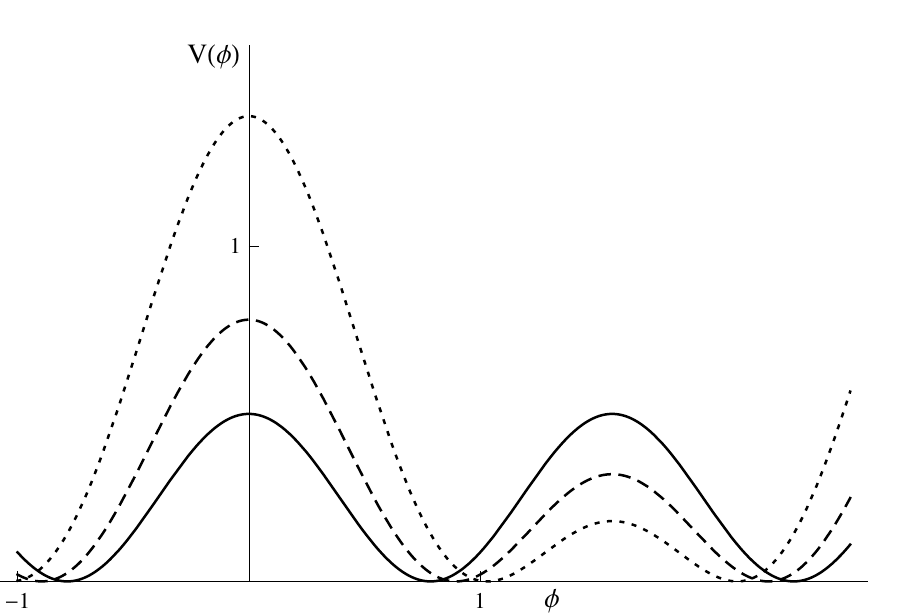}
\caption{The double sine-Gordon model \eqref{2sg} and its two distinct topological sectors, displayed with dotted, dashed and solid lines, for $r=0.6,0.8$ and $1$, respectively.}\label{fig3}
\end{figure}

We now consider a non-polynomial potential, the double sine-Gordon model, which is defined by
\be \label{2sg}
V=\frac1{2r^2}((1+r^2)\cos^2(\phi)-r^2)^2.
\ee
It is displayed in Fig.~\ref{fig3}. This model was studied in \cite{DB} and here we consider $r$ in the interval
$r\in(0,1]$, with the limit $r\to1$ leading us back to the sine-Gordon model already studied; see \eqref{sg}. The first-order equation is
\be
\phi^\prime=\frac1r((1+r^2)\cos^2 (\phi)-r^2).
\ee
There are two distinct kinklike solutions
\bes\ben
&&\phi_{1}(x)=\arctan\left(\frac1r\tanh(x)\right),
\\
&&\phi_{2}(x)={\rm arccot}(r\tanh(x)),
\een\ees
and they can be used to map the other similar topological sectors, as it happens with the sine-Gordon model. They have energies
\bes\ben
&&E_1=2r+2(1-r^2)\arctan(1/r),
\\
&&E_2=2r-2(1-r^2)\arctan(r).
\een\ees
One notes that in the limit $r\to1$ the two solutions $\phi_1$ and $\phi_2$ only differs by a factor of $\pi/2$, so they are similar solutions. Also, in this limit the two energies degenerate to the same unit value $E=1$, as expected.

We use the two solutions to get
\bes\ben
w_1(x)=\frac1{r}((1+r^2)\cosh^2(x)-1),
\\
w_2(x)=\frac1{r}(r^2-(1+r^2)\cosh^2(x)),
\een\ees
which give the two potentials
\bes\label{sgsta}\ben
U_1(x)=\frac12 (1+r^2)g_1(x)\sech^4(x),
\\
U_2(x)=\frac12 (1+r^2)g_2(x)\sech^4(x),
\een\ees
where
\bes\ben
&&g_1(x)=\frac{2(1-r^2)C_2(x)-(1+
r^2)(3-C_4(x))}{\left(r^2+\tanh^2(x)\right)^2},\;\;\;
\\
&&g_2(x)=\frac{-2(1-r^2)C_2(x)-(1+
r^2)(3-\!C_4(x))}{\left(1+r^2\tanh^2(x)\right)^2},\;\;\quad\;\;\;
\een\ees
with $C_2(x)=\cosh(2x)$ and $C_4(x)=\cosh(4x)$. The limit $r\to1$
leads to the potential $4-8\,\sech^2(2x)$, which is the same
potential \eqref{psg} that appeared before in the sine-Gordon
model already studied.

\begin{figure}[t]
\includegraphics[width=7cm]{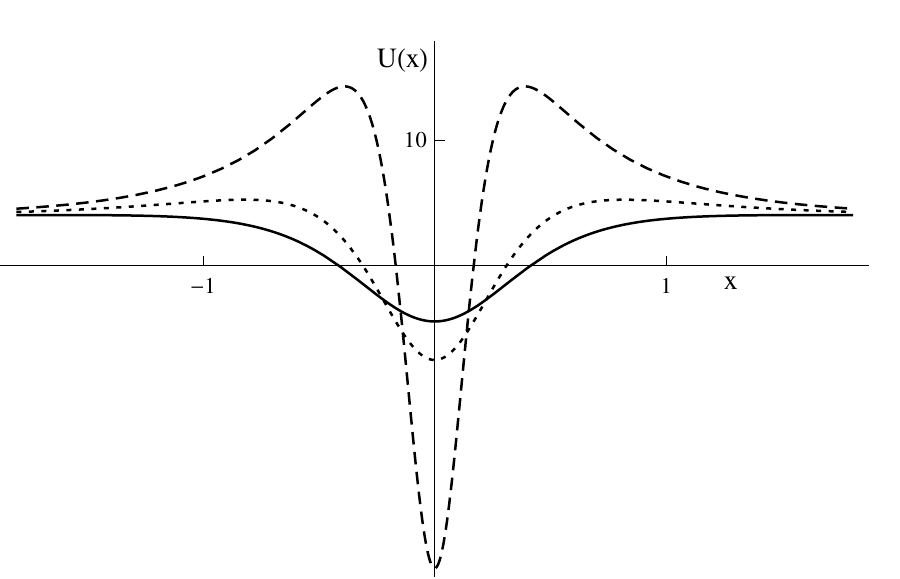}
\includegraphics[width=7cm]{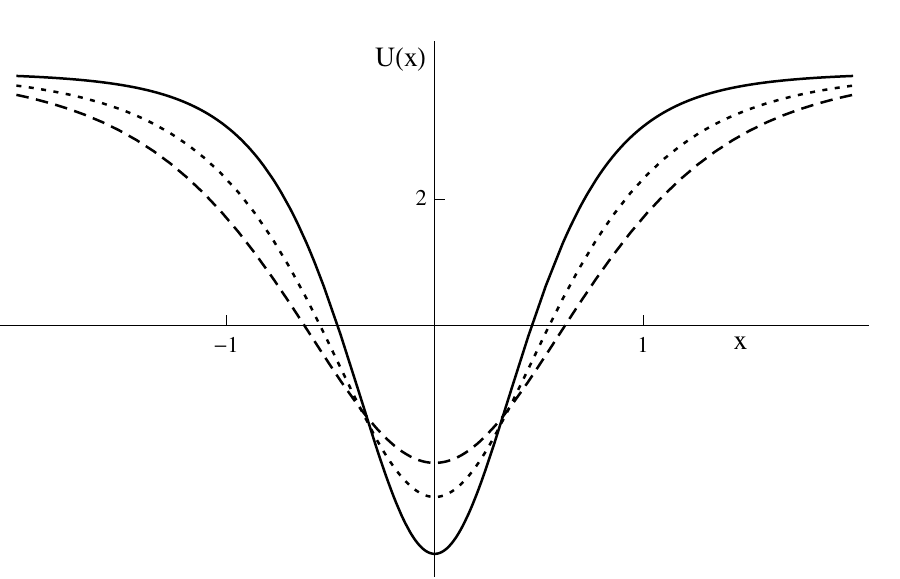}
\caption{The potentials $U_1(x)$ and $U_2(x)$ that appear in
\eqref{sgsta}, displayed in the top and bottom panels, taking
$r=0.3, 0.6,$ and $0.9$, and using dashed, dotted and solid lines,
respectively. }\label{fig4}
\end{figure}

In Fig.~\ref{fig4} we display the two potentials $U_1(x)$ and $U_2(x)$ for some values of $r$.
One notes that the potential $U_1(x)$ changes from the volcano type to the modified P\"oschl-Teller type, as $r$ increases toward unit, while $U_2(x)$ is of the modified P\"oschl-Teller type. However, it is interesting to see that in the limit $r\to1$, both potentials become the same modified P\"oschl-Teller potential.

The above investigations deal with specific scalar field models and are in fact manifestations of a stronger result, which we formulate as:

{\it Consider a standard scalar field theory of the type \eqref{mod} described by first-order differential equations that solve the equation of motion. If it engenders distinct topological sectors with distinct solutions $\phi_i(x), i=1,2,\cdots,$ with distinct energies $E_i, i=1,2,\cdots,$ then there may be distinct stability potentials that can be associated to distinct supersymmetric quantum mechanical systems.}

To see how this assertion works we suppose that the model supports at least two topological sectors, with solutions $\phi_i(x)$ and $\phi_j(x)$ and energies $E_i$ and $E_j$, respectively. We then note that the two sectors induce two stability potentials, which are constructed by the corresponding $f_i(x)$ and $f_j(x)$ as $U_i(x)=f_i^2(x)-f_i^\prime(x)$ and $\alpha^2U_j(\alpha x)=\alpha^2f_j^2(x)-\alpha^2f_j^\prime(\alpha x)$, respectively. Recall that the prime stands for derivative with respect to the argument.  We now suppose that they describe equivalent systems, such that
\be\label{ff}
(f_i(x)\!+\!\alpha f_j(\alpha x))(f_i(x)\!-\!\alpha f_j(\alpha x))\!=\!\frac{d}{dx}(f_i(x)\!-\!\alpha f_j(\alpha x)).
\ee
A possibility to validate this expression is to make $f_i(x)=\alpha f_j(\alpha x)$, and here one gets $w_i(x)=c\,w_j(\alpha x)$, where $c$ is a real constant. This imposes that the energies obey $E_j=|\alpha|c^2E_i$. We can then make $c^2=1=\alpha^2$ to get $E_i=E_j$. One notes that the case $c=1$ and $\alpha=-1$ leads to $U_i(x)=U_j(-x)$, as it occurs in the $\phi^6$ model and also in the $\phi^8$ model, in the case of the two lateral sectors $U_1(x)$ and $U_3(x)=U_1(-x)$. Also, the case $c=\pm1$ and $\alpha=1$ leads to $U_i(x)=U_j(x)$, as it is in the sine-Gordon model.

We can also make $E_i=E_j$ with $c^2=1/|\alpha|\neq 1$. In this case, however, it can be shown that if $\phi_i$ solves the first-order equation,
$\phi_j$ does not. Moreover, one can also take $c^2|\alpha|\neq 1$ to make $E_i\neq E_j$, but here the same reasoning applies, that is, if $\phi_i$ solves the first-order equation, $\phi_j$ does not. Thus, the equality $w_i(x)=c\,w_j(\alpha x)$ only works if the two sectors have the same energy and $c^2=1=\alpha^2$.

On the other hand, if $f_i(x)\neq \alpha f_j(\alpha x)$, there is still another possibility to validate \eqref{ff} to make the two stability potentials equivalent; it demands that $E_i\neq E_j$ and
\be\label{ww}
\frac{d}{dx}\left(\frac{w_i(x)}{w_j(\alpha x)}\right)=\kappa\, w_i^2(x),
\ee
where $\kappa$ is an integration constant. If this is not satisfied, one is then left with distinct potentials, as we have illustrated with the
$\phi^8$ model, and also with the double sine-Gordon model. We tried but we could not find a scalar field model of the type studied in this work, that could serve to fulfill the above expression \eqref{ww}.

\section{Ending comments}
\label{sec4}

In this work we studied the route that starts from a scalar field theory and ends on quantum mechanics. We illustrated the investigation with some specific examples, and then described how a scalar field theory may be defined to lead to distinct stability potentials, which may be used as supersymmetric quantum mechanical problems. We believe that the results of this work will stimulate new investigations on the subject. In particular, we are now searching for more general models, that support three or more distinct topological sectors. Also, we are studying the inverse route, the route in which one starts from quantum mechanics to reconstruct the scalar field theory. We hope to report on this in the near future.

\acknowledgments{The authors would like to thank Laercio Losano for comments, and the Brazilian agencies CAPES and CNPq for financial support. DB also thanks support from the CNPq grants 455931/2014-3 and 306614/2014-6.}

\end{document}